\documentclass[journal]{IEEEtran}

\usepackage{xcolor,soul,framed} 

\colorlet{shadecolor}{yellow}
\usepackage[pdftex]{graphicx}
\usepackage{graphicx}
\usepackage{svg}
\graphicspath{{../pdf/}{../jpeg/}{../emf}{../svg}}
\DeclareGraphicsExtensions{.pdf,.jpeg,.png,.emf,.svg}

\usepackage[cmex10]{amsmath}
\usepackage{array}
\usepackage{mdwmath}
\usepackage{mdwtab}
\usepackage{eqparbox}
\usepackage{url}
\usepackage{float}
\usepackage{tikz}
\usepackage{amssymb}
\usepackage{amstext}

\usepackage[colorlinks=true, linkcolor=blue, citecolor=blue, urlcolor=blue]{hyperref}

\usepackage{orcidlink}

\usetikzlibrary{arrows.meta, positioning}

\begin{document}
\bstctlcite{IEEEexample:BSTcontrol}
    \title{Deep Reinforcement Learning-Aided Frequency Control of LCC-S Resonant Converters for Wireless Power Transfer Systems}
  \author{Reza Safari\orcidlink{0000-0002-1489-4055},
      Mohsen Hamzeh\orcidlink{0000-0001-6919-9706},~\IEEEmembership{Member,~IEEE,}\\
      and~Nima Mahdian Dehkordi\orcidlink{0000-0003-1493-1941}
    \thanks{Reza Safari and Mohsen Hamzeh are with the Department of Electrical Engineering, University of Tehran, Tehran, Iran (emails: safari.reza@ut.ac.ir, mohsenhamzeh@ut.ac.ir).}%
    \thanks{Nima Mahdian Dehkordi is with the Department of Electrical Engineering, Shahid Rajaee Teacher Training University, Tehran, Iran (email: nimamahdian@sru.ac.ir).}%
    }  
\markboth{IEEE TRANSACTIONS ON INDUSTRIAL ELECTRONICS
}{Roberg \MakeLowercase{\textit{et al.}}: High-Efficiency Diode and Transistor Rectifiers}

\maketitle

\begin{abstract}
This paper presents a novel deep reinforcement learning (DRL)-based control strategy for achieving precise and robust output voltage regulation in LCC-S resonant converters, specifically designed for wireless power transfer applications. Unlike conventional methods that rely on manually tuned PI controllers or heuristic tuning approaches, our method leverages the Twin Delayed Deep Deterministic Policy Gradient (TD3) algorithm to systematically optimize PI controller parameters. The complex converter dynamics are captured using the Direct Piecewise Affine (DPWA) modeling technique, providing a structured approach to handling its nonlinearities. This integration not only eliminates the need for manual tuning, but also enhances control adaptability under varying operating conditions. The simulation and experimental results confirm that the proposed DRL-based tuning approach significantly outperforms traditional methods in terms of stability, robustness, and response time. This work demonstrates the potential of DRL in power electronic control, offering a scalable and data-driven alternative to conventional controller design approaches.
\end{abstract}


%
\IEEEpeerreviewmaketitle


\section{Introduction}

\IEEEPARstart{O}{ver} the past few decades, WPT has become a game-changing technology that is attracting more and more attention worldwide. EVs, medicinal implants, and consumer electronics are just a few of the industries that have benefited from WPT's ability to transmit electrical energy without the need for physical connectors. Continuous research efforts to increase efficiency, range, and adaptability have been spurred by its capacity to improve convenience, safety, and system reliability. In order to improve WPT technology, researchers are constantly investigating new circuit topologies, control schemes, and optimization approaches in response to the growing need for effective energy transmission solutions
\cite{Power_Stability_Enhancement_Inductive_Wireless_Power_Transfer_Systems_A_Review}.

Several DC-DC converter topologies for WPT systems are examined in \cite{Comprehensive_Study_WPT_Topologies}, with an emphasis on both high-power and low-power applications. Through switched-capacitor correction, a 150 W single-stage converter in \cite{WPT_Topologies_1} achieves great efficiency, guaranteeing steady performance even in the presence of misalignment. A 1 kW single-phase converter with 94.35\% efficiency and constant frequency operation is introduced by \cite{WPT_Topologies_2} for low-power systems. On the other hand, \cite{WPT_Topologies_3} offers a 50 kW bidirectional converter with an efficiency of 95.4\% for high-power applications. Key enhancements that address the drawbacks of multi-phase converters include decreased complexity, improved stability, and maximized efficiency throughout load ranges \cite{Comprehensive_Study_WPT_Topologies,WPT_Topologies_3}. In order to provide high efficiency and consistent output under a variety of situations, \cite{LCCS_Ramezani} suggests an optimized LCC-S resonant converter for stationary EV wireless charging. It reaches a maximum efficiency of 94.8\% by using zero-voltage switching (ZVS) and a time-weighted optimization approach. Its efficacy is validated using a 1 kW prototype.

Analyzing the internal dynamics of converters, as inherently non-linear systems \cite{small_signal_model_dcm}, is crucial, regardless of their architecture and any accompanying benefits or drawbacks. In order to analyze transient behaviors, ensure robust control, and maximize efficiency under a variety of operating situations, accurate dynamic modeling is essential \cite{Dynamics_issue}.

The typical approaches to capture low-frequency behavior in power electronic converters (PECs) include small-signal modeling techniques \cite{switched_2} or average PEC models \cite{switched_1}. These methods provide a macroscopic representation of the system by transforming discontinuous models into continuous ones \cite{switched_10}. They employ methods such as state-space averaging of converter state variables or average switch modeling of terminal waveforms, which ignore high-frequency switching dynamics \cite{review_small_signal_large_signal}.

Unlike average models, switched models are the most accurate for studying dynamic stability issues since they take into account all nonlinearities \cite{review_small_signal_large_signal}. Piecewise affine (PWA) approaches stand out among switched models. By breaking up a switching period into multiple subintervals, these techniques greatly improve model accuracy \cite{switched_4}. Additional progress in this area may be seen in studies like \cite{molla2014hybrid_switched_5,switched_6}, which use switched affine models to investigate series resonant converters (SRCs). Furthermore, research such as \cite{switched_8} shows that state-plane-based techniques can capture system dynamics more effectively than traditional methods \cite{switched_9}.

Effective control is becoming more and more challenging due to issues like the complexity of models for controlled objects in PECs. Real-time and accurate information about the controlled object is frequently the foundation of traditional PEC control systems \cite{rl_review}. A promising method for addressing perceptual decision-making difficulties in complex systems is DRL \cite{rl_1, rl_2}. DRL control strategies work with simulation models or real-world systems to create policies. By gradually accumulating experience, these goal-oriented tactics teach a policy to operate at its best. In contrast to conventional techniques, DRL eliminates offline learning based on predetermined data sets and does not require explicit models or comprehensive environmental knowledge \cite{rl_3}.

Recent research extends the success of DRLs to diverse power electronic systems, enhancing stability and adaptability. Deep Q learning networks (DQN) were investigated in \cite{rl_35} for voltage regulation, and their performance was compared to traditional methods. DDPG-based approaches were applied in \cite{rl_36, rl_2} and \cite{rl_37}, with techniques such as ADRC, PI, and sliding mode observers (SMO) being integrated to enhance robustness and stability. PPO combined with traditional control strategies was explored for improved voltage control in buck-boost converters \cite{rl_38}. For optimal modulation in dual active bridge (DAB) converters, Q-learning and twin delayed deep-deterministic (TD3) were utilized with additional enhancements such as non-linear disturbance observers and backstepping controllers being incorporated \cite{rl_6,rl_61}. Through these studies, the effectiveness of DRL techniques in improving stability and adaptability in DC/DC converter control systems has been highlighted.


In WPT applications, the internal dynamics of LCC-S resonant converters has not been systematically analyzed, particularly for frequency-based control. Additionally, existing WPT applications rely on manually tuned PI controllers, which lack robustness under varying conditions. To date, no systematic approach has been proposed for tuning PI parameters when frequency serves as the control input, limiting control optimization and stability.

In this work, the research gap is addressed by employing DPWA modeling in combination with DRL for varied-frequency control of the LCC-S resonant converter. A PI controller is designed and fine-tuned using the TD3 algorithm, enabling dynamic optimization of system performance and ensuring robust control under varying operating conditions. The effectiveness of the proposed method is demonstrated through simulation and experimental results, showcasing significant improvements in control accuracy and robustness compared to traditional tuning techniques.


\section{Switched Systems Model}

Switched systems are an essential class of hybrid systems characterized by transitions between multiple subsystems, making them highly versatile for various control and optimization tasks. In this work, we focus on one prominent switched system model: DPWA model which is an extension of the PWA model \cite{molla2014hybrid_switched_5}.
A graphical representation of DPWA systems with polyhedral cells is shown in Fig. 1, where tunable boundaries enable non-autonomous switching, and non-tunable boundaries result in autonomous switching.

\begin{figure}[b]
  \begin{center}
  \includegraphics[width=2.5in]{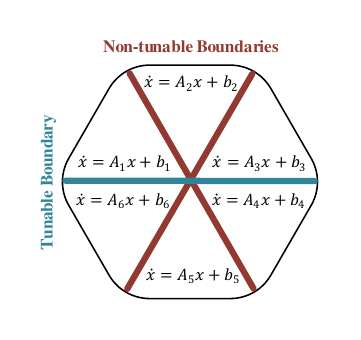}\\
  \vspace{-25pt}
  \caption{A basic visual diagram of DPWA systems with six polyhedral regions.}\label{DPWA_Poly}
  \end{center}
\end{figure}

\begin{figure}[b]
  \begin{center}
  \includegraphics[width=3.7in]{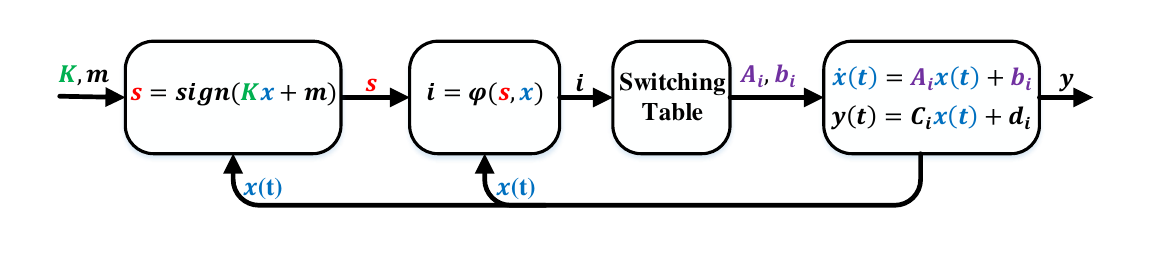}\\
  \vspace{-15pt}
  \caption{Block diagram of a resonant converter using DPWA technique \cite{molla2014hybrid_switched_5}.}\label{DPWA_Poly}
  \end{center}
\end{figure}

The DPWA model introduces tunable cell boundaries, making it ideal for applications requiring adaptive switching, such as in PECs like resonant converters. The DPWA model is formulated as
\begin{equation}
\begin{cases}
\dot{x} = A_i x + b_i \\
y = C_i x + d_i
\end{cases}, \quad i = 1, 2, \ldots, M
\end{equation}

here, the subscript $i$ denotes the cell in which the state trajectory resides. Cell boundaries can be either polyhedral or ellipsoidal. $M$ represents the number of subsystems, while $A_i$, $b_i$, $C_i$, and $d_i$ are matrices characterizing the dynamics of the $i$-th subsystem.

Fig. 2 illustrates the DPWA method for modeling the converter, whose stability can be analyzed as an LMI optimization problem \cite{molla2014hybrid_switched_5}. A feasible solution to this problem guarantees global asymptotic stability and allows for the calculation of the controller input $K$. Conversely, the existence of $K$ implies a feasible solution and system stability. This relationship directly links variations in $K$ to variations in $s$. Thus, tunable boundaries can be easily controlled by varying the parameter $K$.
Hence, to systematically analyze stability and design a controller, the switching command is defined as follows:
\begin{equation}
s = \mathrm{sign}(\mathbf{K}x + \mathbf{m}), \quad x \in \mathbb{R}^n
\end{equation}

\section{The DPWA Model of LCC-S Converter}
Fig. 3 illustrates the configuration of a LCC-S resonant converter. The proposed converter utilizes a resonant tank on the primary side, consisting of an inductor ($L_1$) and two capacitors ($C_1$ and $C_{s1}$). The magnetic coupling between the primary and secondary sides is represented by equivalent inductors ($L_p$, $L_s$, and $M$). On the secondary side, a resonant tank includes a series capacitor ($C_{s2}$), while a parallel combination of a capacitor ($C_I$) and a resistor ($R_I$) models the output filter and load, respectively. Furthermore,  in this schematic, the variables $x_1$, $x_2$, and $x_3$ represent the current through the tank inductor, the voltage across the first capacitor on the primary side of the tank, and the voltage across the second capacitor on the primary side of the tank, respectively. Additionally, $x_4$ and $x_5$ denote the input and output currents of the transformer, while $x_6$ corresponds to the voltage across the secondary side network's capacitor, and $x_7$ signifies the output voltage. The variable $V_s$ also represents the output voltage of the full-bridge.

\begin{figure}[t]
 \raggedleft
 \centering
  \hspace{-40pt} 
 \rotatebox{0}{\includegraphics[width=4.0in]{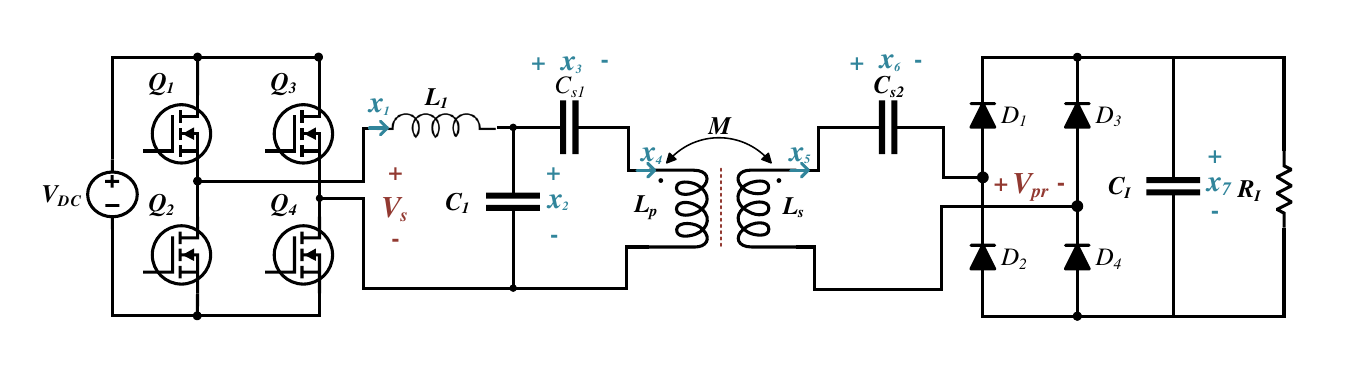}}
 \vspace{-20pt}
 \caption{LCC-S resonant converter.}\label{LCC_S}
\end{figure}

\begin{figure}[t]
  \begin{center}
  \includegraphics[width=3.5in]{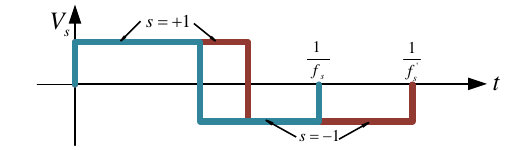}\\
  \caption{The variation of the voltage $V_s$ under frequency control method.}\label{DPWA_Poly}
  \end{center}
\end{figure}

By considering the frequency control strategy shown in Fig. 4, the LCC-S resonant converter inherently operates as a switched system with affine subsystems. Based on the states of the switches and the output rectifier, six distinct modes or subsystems can occur. These modes are illustrated in Fig. 5 under various conditions. These subintervals are primarily determined by the switching states of transistors $Q_1$, $Q_2$, $Q_3$, $Q_4$ and diodes $D_1$, $D_2$, $D_3$, and $D_4$. The behavior of each state variable and their interactions influence the converter's operating mode.

In this context, $V_f$ represents the forward voltage drop across the output diodes, while $R_1$, $r_p$ and $r_s$ correspond to the combined resistance of the inductor and the series resistance of the capacitors. The variable $s \in \{-1, +1\}$ is a logic input that determines the state of the input switches. $s = \pm 1$ \text{ switches between two configurations: } $Q_1$, $Q_4$ \text{ON or } $Q_2$, $Q_3$ \text{ON}. Notably, there is no direct relationship between the operating frequency and the state variables of the system.

\begin{figure}[t]
  \begin{center}
  \includegraphics[width=3.5in]{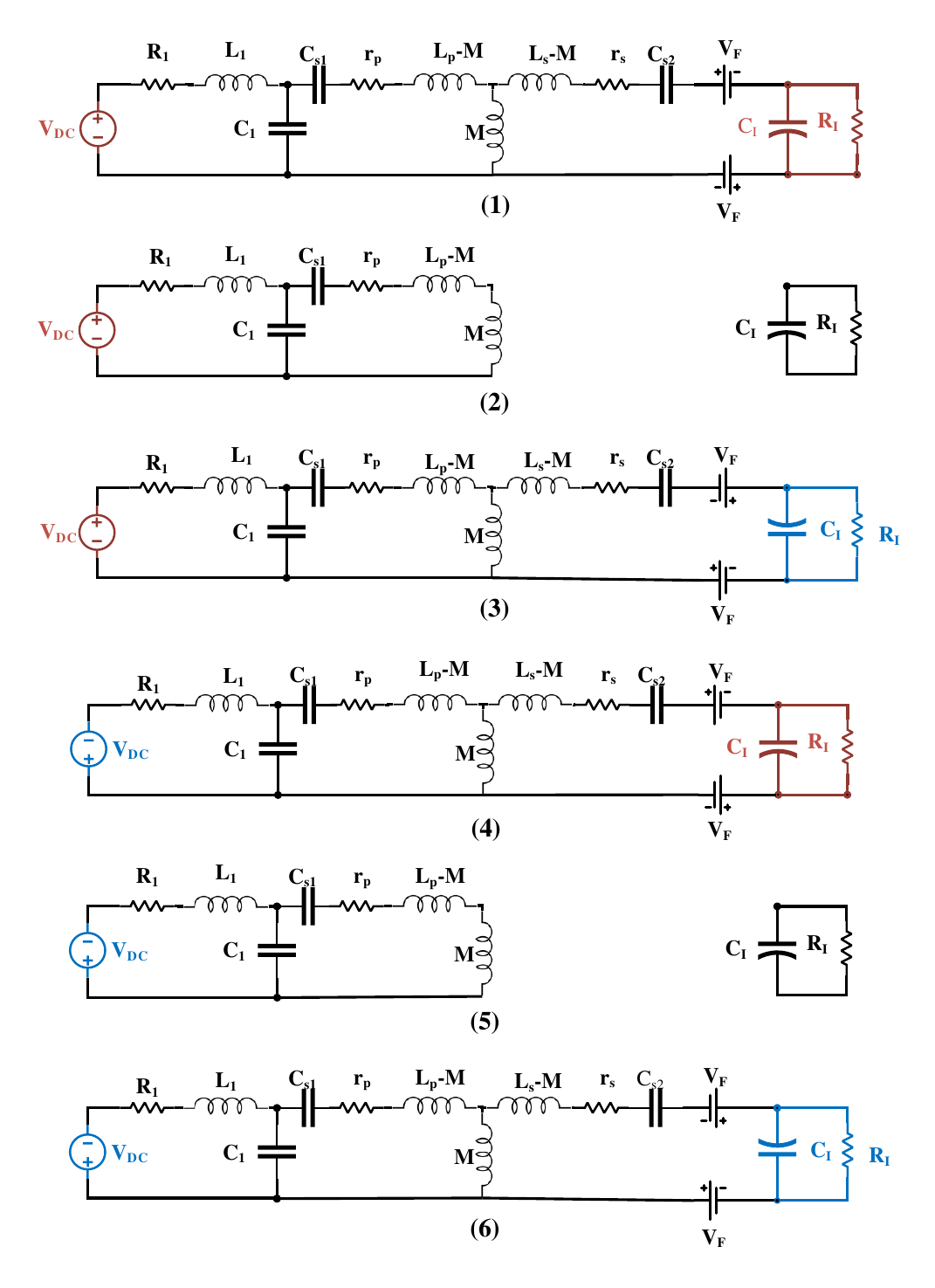}\\
  \vspace{-18pt}
  \caption{The six modes of LCC-S converter with following conditions as: (1) $s=1$, $x_5>x_4$, (2) $s = 1$, $|v_{pr}|<x_7$, (3) $s = 1$, $x_5<x_4$, (4) $s=-1$, $x_5<x_4$, (5) $s=-1$, $|v_{pr}|<x_7$, (6) $s=-1$, $x_5>x_4$ .}\label{DPWA_Poly}
  \end{center}
\end{figure}

The logic input $s$ is defined as follows to convert the constrained switched affine model into a DPWA model:
\begin{equation}
s = \begin{cases}
+1 & Kx + m \geq 0 \\
-1 & Kx + m < 0
\end{cases}
\end{equation}
here, $K$ = [$k_1$, $k_2$, $k_3$, $k_4$, $k_5$, $k_6$, $k_7$] and $m$ represent the parameters of the switching surfaces that determine the variable $s$.

The DPWA representation of the LCC-S model is expressed as:
\setlength{\parindent}{0pt}
\begin{equation}
\begin{cases}
\dot{x}(t) = A_i x(t) + b_i \\
y(t) = Cx(t) \\
i = \varphi(s,x(t))
\end{cases}
\end{equation}

\begin{table*}[h!]
\centering
\caption{Mathematical Representation of the LCC-S Resonant Converter Using State-Space Matrices}
\label{tab:subsystems}
\begin{tabular}{c >{\centering\arraybackslash}m{11cm} >{\centering\arraybackslash}m{2cm}}
\hline
$\centering i$ & \centering ${A_i} $& $\centering {b_i}$ \\
\hline
1 & $\centering \begin{pmatrix}
\frac{-R_1}{L_1} & \frac{-1}{L_1} & 	0 	  & 		0	   & 0 & 0 & 0     \\
\frac{1}{C_1}    &        0       & 	0     & \frac{-1}{C_1} & 0 & 0 & 0     \\
	0			 &        0       & 	0     & \frac{-1}{C_{s1}} & 0 & 0 & 0  \\
	0			 & \frac{L_s' + M}{L_{eq}}    & \frac{L_s' + M}{L_{eq}}  & \frac{-r_p(L_s' + M)}{L_{eq}}  & \frac{-r_sM}{L_{eq}} & \frac{-M}{L_{eq}} & \frac{-M}{L_{eq}}   \\
	0			 & \frac{M}{L_{eq}}    & \frac{M}{L_{eq}}  & \frac{-r_pM}{L_{eq}}  & \frac{-r_s(L_p'+M)}{L_{eq}} & \frac{-(L_p'+M)}{L_{eq}} & \frac{-(L_p'+M)}{L_{eq}}   \\
    0            &      0              &         0         &         0        & \frac{-1}{C_{s2}}  &      0      &       0    \\   
    0            &      0              &         0         &         0        &	\frac{1}{C_{1}}  &      0      &       \frac{-1}{R_1C_1}    \\
\end{pmatrix}$ & $\centering \begin{bmatrix}
\frac{V_{in}}{L_1} \\
0 \\
0 \\
\frac{-2MV_F}{L_{eq}} \\
\frac{-2(L_p' + M)V_F}{L_{eq}} \\
0 \\
0 \\
\end{bmatrix}$\\
\hline
2 & $\begin{pmatrix}
\frac{-R_1}{L_1} & \frac{1}{L_1} & 0 & 0 & 0 & 0 & 0 \\
\frac{1}{C_1} & 0 & 0 & \frac{1}{C_1} & 0 & 0 & 0 \\
0 & 0 & 0 & \frac{-1}{C_{s1}} & 0 & 0 & 0 \\
0 & \frac{1}{L_p'+M} & \frac{1}{L_p'+M} & \frac{-r_p}{L_p'+M} & 0 & 0 & 0 \\
0 & 0 & 0 & 0 & 0 & 0 & 0 \\
0 & 0 & 0 & 0 & 0 & 0 & 0 \\
0 & 0 & 0 & 0 & 0 & 0 & \frac{-1}{R_1C_1} \\
\end{pmatrix}$ & 
$\begin{bmatrix}
\frac{V_{in}}{L_1} \\
0 \\
0 \\
0 \\
0 \\
0 \\
0 \\
\end{bmatrix}$ \\
\hline
3 &$ \centering \begin{pmatrix}
\frac{-R_1}{L_1} & \frac{1}{L_1} & 	0 	  & 		0	   & 0 & 0 & 0     \\
\frac{1}{C_1}    &        0       & 	0     & \frac{1}{C_1} & 0 & 0 & 0     \\
	0			 &        0       & 	0     & \frac{-1}{C_{s1}} & 0 & 0 & 0  \\
	0			 & \frac{L_s' + M}{L_{eq}}    & \frac{-(L_s' + M)}{L_{eq}}  & \frac{-r_p(L_s' + M)}{L_{eq}}  & \frac{r_sM}{L_{eq}} & \frac{M}{L_{eq}} & \frac{M}{L_{eq}}   \\
	0			 & \frac{-M}{L_{eq}}    & \frac{-M}{L_{eq}}  & \frac{r_pM}{L_{eq}}  & \frac{-r_s(L_p'+M)}{L_{eq}} & \frac{-(L_p'+M)}{L_{eq}} & \frac{-(L_p'+M)}{L_{eq}}   \\
    0            &      0              &         0         &         0        & \frac{1}{C_{s2}}  &      0      &       0    \\   
    0            &      0              &         0         &         0        &	\frac{1}{C_{1}}  &      0      &       \frac{-1}{R_1C_1}    \\
\end{pmatrix}$ &$ \centering \begin{bmatrix}
\frac{-V_{in}}{L_1} \\
0 \\
0 \\
\frac{2MV_F}{L_{eq}} \\
\frac{-2(L_p' + M)V_F}{L_{eq}} \\
0 \\
0 \\
\end{bmatrix}$ \\
\hline
4 & $\centering \begin{pmatrix}
\frac{-R_1}{L_1} & \frac{-1}{L_1} & 	0 	  & 		0	   & 0 & 0 & 0     \\
\frac{1}{C_1}    &        0       & 	0     & \frac{-1}{C_1} & 0 & 0 & 0     \\
	0			 &        0       & 	0     & \frac{-1}{C_{s1}} & 0 & 0 & 0  \\
	0			 & \frac{L_s' + M}{L_{eq}}    & \frac{L_s' + M}{L_{eq}}  & \frac{-r_p(L_s' + M)}{L_{eq}}  & \frac{-r_sM}{L_{eq}} & \frac{-M}{L_{eq}} & \frac{-M}{L_{eq}}   \\
	0			 & \frac{M}{L_{eq}}    & \frac{M}{L_{eq}}  & \frac{-r_pM}{L_{eq}}  & \frac{-r_s(L_p'+M)}{L_{eq}} & \frac{-(L_p'+M)}{L_{eq}} & \frac{-(L_p'+M)}{L_{eq}}   \\
    0            &      0              &         0         &         0        & \frac{1}{C_{s2}}  &      0      &       0    \\   
    0            &      0              &         0         &         0        &	\frac{1}{C_{1}}  &      0      &       \frac{-1}{R_1C_1}    \\
\end{pmatrix}$ & $\centering \begin{bmatrix}
\frac{-V_{in}}{L_1} \\
0 \\
0 \\
\frac{-2MV_F}{L_{eq}} \\
\frac{-2(L_p' + M)V_F}{L_{eq}} \\
0 \\
0 \\
\end{bmatrix} $\\
\hline
5 & $\begin{pmatrix}
\frac{-R_1}{L_1} & \frac{-1}{L_1} & 0 & 0 & 0 & 0 & 0 \\
\frac{1}{C_1} & 0 & 0 & \frac{-1}{C_1} & 0 & 0 & 0 \\
0 & 0 & 0 & \frac{-1}{C_{s1}} & 0 & 0 & 0 \\
0 & \frac{1}{L_p'+M} & \frac{1}{L_p'+M} & \frac{-r_p}{L_p'+M} & 0 & 0 & 0 \\
0 & 0 & 0 & 0 & 0 & 0 & 0 \\
0 & 0 & 0 & 0 & 0 & 0 & 0 \\
0 & 0 & 0 & 0 & 0 & 0 & \frac{-1}{R_1C_1} \\
\end{pmatrix}$ & 
$\begin{bmatrix}
\frac{-V_{in}}{L_1} \\
0 \\
0 \\
0 \\
0 \\
0 \\
0 \\
\end{bmatrix}$ \\
\hline
6 & $\begin{pmatrix}
\frac{-R_1}{L_1} & \frac{-1}{L_1} & 0 & 0 & 0 & 0 & 0 \\
\frac{1}{C_1} & 0 & 0 & \frac{1}{C_1} & 0 & 0 & 0 \\
0 & 0 & 0 & \frac{-1}{C_{s1}} & 0 & 0 & 0 \\
0 & \frac{L_s' + M}{L_{eq}} & \frac{-(L_s' + M)}{L_{eq}} & \frac{-r_p(L_s' + M)}{L_{eq}} & \frac{r_sM}{L_{eq}} & \frac{M}{L_{eq}} & \frac{M}{L_{eq}} \\
0 & \frac{-M}{L_{eq}} & \frac{-M}{L_{eq}} & \frac{r_pM}{L_{eq}} & \frac{-r_s(L_p'+M)}{L_{eq}} & \frac{-(L_p'+M)}{L_{eq}} & \frac{-(L_p'+M)}{L_{eq}} \\
0 & 0 & 0 & 0 & \frac{1}{C_{s2}} & 0 & 0 \\
0 & 0 & 0 & 0 & \frac{1}{C_{1}} & 0 & \frac{-1}{R_1C_1}
\end{pmatrix}$ & 
$\begin{bmatrix}
\frac{V_{in}}{L_1} \\
0 \\
0 \\
\frac{2MV_F}{L_{eq}} \\
\frac{-2(L_p' + M)V_F}{L_{eq}} \\
0 \\
0
\end{bmatrix}$ \\
\hline
\end{tabular}
\end{table*}

\hspace{10pt} In this case, $A_i$ and $b_i$ are listed in Table I, and considering $L_{eq}$ = $L_p'$$L_s'$ + $M$($L_p'$+$L_s'$) with $C = [0 0 0 0 0 0 1]$ (where $x_7$ is the output of the converter), and condition function is defined as:
\begin{equation}
\varphi(s, x) = \begin{cases}
1 &   s = +1, \quad x_5 > x_4 \\
2 &   s = +1, \quad |v_{pr}| < x_7 \\
3 &   s = +1, \quad x_5 < x_4 \\
4 &   s = -1, \quad x_5 < x_4 \\
5 &   s = -1, \quad |v_{pr}| < x_7 \\
6 &   s = -1, \quad x_5 > x_4 \\
\end{cases}
\end{equation}

The input voltage of the rectifier, denoted as $v_{pr}$, is given by the following equation:
\begin{equation}
v_{pr} = \displaystyle \frac{M}{L_p' + M}(x_2 + x_3 - r_px_4) - r_sx_5 - x_6
\end{equation}

\hspace{10pt} A DPWA model of a full-bridge DC-DC LCC-S resonant converter was simulated in MATLAB/SIMULINK to evaluate the accuracy of the model. The components of the converter are listed in Table II \cite{LCCS_Ramezani}. Based on these components, the main resonant frequency was set to 85 kHz, which falls within the standard operating frequency range of 79 to 90 kHz for WPT systems. This frequency was deliberately chosen, as it represents the primary frequency typically used in WPT under standard conditions \cite{SAEJ2954_2019}.

\hspace{10pt} For the DPWA model, the control parameters were set as $K=[5413.4, 265.6, 463.1, 1445.4, 370.1, -1]$ and $m=2208.7$, corresponding to a switching frequency of $f_s = 85 \text{ kHz}$. The simulation results for this case are shown in Fig. 6. The results clearly demonstrate that the model operates correctly and accurately predicts the behavior of the state-space variables at the main resonant frequency. Furthermore, the six distinct subsystems identified during the DPWA modeling process are clearly distinguishable and validated under specific operating conditions.

\hspace{10pt} Using the DPWA model, as illustrated in Fig. 2, variations in $K$ result in changes to the logic input $s$. Consequently, no external controller is required, as the state of the switches automatically changes at the boundaries of a hyperplane defined by $Kx + m = 0$. Therefore, adjusting $K$ is equivalent to varying the switching frequency.

\hspace{10pt} To achieve robust voltage regulation, a controller is designed to monitor the error between the desired and actual output voltages and adjust $K$ accordingly. However, the inherent complexity and non-linearity of the system make designing a sophisticated controller challenging. To maintain simplicity, a basic proportional-integral (PI) controller is employed. The subsequent section will explore the design and tuning of this PI controller.

\begin{table}[ht]
\caption{Components of the LCC-S Converter and Their Values}
\label{tab:components}
\centering
\begin{tabular}{|c|c|c|}
\hline
Symbol & Parameter & Value \\
\hline
$L_1$ & Ser. Pri. Ind. $(\mu$H) & 73.4 \\
\hline
$C_1$ & Par. Pri. Cap. (nF) & 46.7 \\
\hline
$C_{s1}$ & Ser. Pri. Cap. (nF) & 15.5 \\
\hline
$C_{s2}$ & Ser. Sec. Cap. (nF) & 11.7 \\
\hline
$R_1$ & Eq. Res. Ser. Pri. Ind. (m$\Omega)$ & 50 \\
\hline
$r_p$ & Eq. Res. Cou. Pri. Ind. (m$\Omega)$ & 382 \\
\hline
$r_s$ & Eq. Res. Cou. Sec. Ind. (m$\Omega)$ & 394 \\
\hline
$V_{dc}$ & Input Volt. (V) & 200 \\
\hline
$L_p$ & Cou. Pri. Ind. $(\mu$H) & 281.3 \\
\hline
$L_s$ & Cou. Sec. Ind. $(\mu$H) & 278.3 \\
\hline
$k$ & Cou. Coeff. & 0.3 \\
\hline
$V_F$ & For. Volt. Drop Diode (V) & 0.7 \\
\hline
$C_I$ & Out. Filt. Cap. $(\mu$F) & 300 \\
\hline
$R_I$ & Out. Res. $(\Omega)$ & 44.77 \\
\hline
\end{tabular}
\end{table}

\begin{figure}[ht]
  \begin{center}
  \includegraphics[width=3.4in]{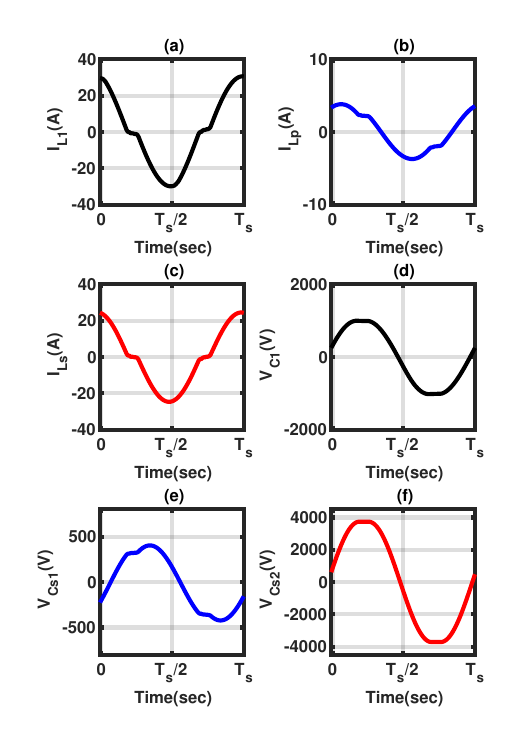}\\
  \vspace{-25pt}
  \caption{Characteristic waveforms of state-space variables: (a) Inductor current \( I_{L1} \), (b) Inductor current \( I_{Lp} \), (c) Inductor current \( I_{Ls} \), (d) Capacitor voltage \( V_{C1} \), (e) Capacitor voltage \( V_{Cs1} \), (f) Capacitor voltage \( V_{Cs2} \).}\label{DPWA_Poly}
  \end{center}
\end{figure}

\section{Tuning PI Controller for LCC-S Converter}
\hspace{10pt} The deep deterministic policy gradient (DDPG) technique is one of the most popular DRL algorithms; nevertheless, it is sensitive to hyperparameters and frequently overestimates Q values obtained from mimicked Q functions, which may render the learned policy erroneous \cite{rl_4}. The TD3 technique \cite{td3_ref} was developed with three significant enhancements to address these problems: 1. Reduced overestimation bias by clipped double Q-learning; 2. Improved stability through delayed policy updates; and 3. Reduced detrimental policy variations through regularization of the smoothing of the target policy \cite{td3_ref}. These enhancements enable the TD3 algorithm to achieve superior performance and faster learning compared to DDPG \cite{rl_6, td3_ref}.


\begin{figure}[t]
  \begin{center}
  \includegraphics[width=3.5in]{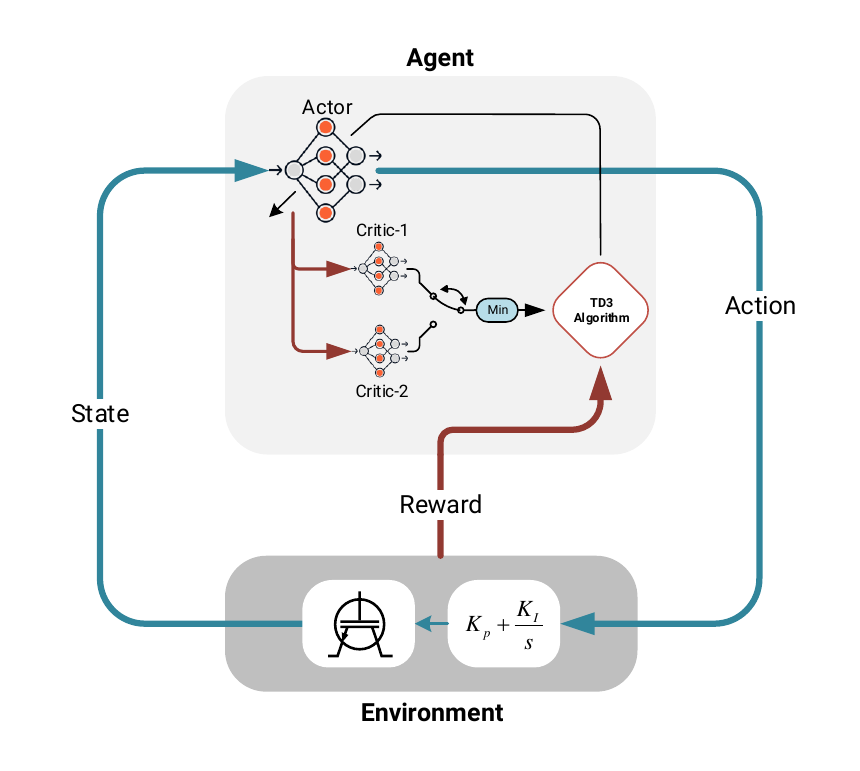}\\
  \vspace{-15pt}
  \caption{A DRL-based approach for tuning the PI controller of a LCC-S converter.}\label{DPWA_Poly}
  \end{center}
\end{figure}
  \vspace{-8pt}
\subsection {Theoretical Foundations of the TD3 Algorithm}
\hspace{10pt} TD3 is an extension of the DDPG algorithm that overcomes its several key limitations. To mitigate overestimation bias and improve stability, TD3 employs two critic networks to independently evaluate the same action. By comparing the estimates of both critics, TD3 reduces the risk of overestimating the potential rewards of an action. Furthermore, TD3 introduces delayed policy updates, meaning that the actor network (which controls the actions) is updated less frequently than the critic networks. This decoupling helps stabilize the learning process and prevents the policy from becoming too sensitive to short-term fluctuations in estimated rewards. Finally, TD3 incorporates target noise, a technique that adds random noise to target actions used to update the critic networks. This smoothing effect helps prevent the policy from exploiting minor inaccuracies in the value estimates, leading to more robust and stable learning. Fig. 7 visually represents the TD3 process, highlighting the use of two critical networks.
The core idea behind TD3 is to minimize the following loss function:
\begin{equation}
L_{k}=\frac{1}{2 M} \sum_{i=1}^{M}\left(y_{i}-Q_{k}\left(S_{i}, A_{i} ; \phi_{k}\right)\right)^{2}
\end{equation}
where, $L_k$ is the loss function for critic $k$, $M$ is the mini-batch size, $y_i$ is the value function target, $Q_k$ is the critic function, $S_i$ is the current observation, $A_i$ is the current action and $\phi_k$ is the critic parameters.

\hspace{10pt} The actor network is updated to maximize the expected return, which can be approximated using the following gradient.
\begin{equation}
\nabla_{\theta} J \approx \frac{1}{M} \sum_{i=1}^{M} G_{a_{i}} G_{\pi_{i}}
\end{equation}
where, $G_{a_i}$ is the gradient of the minimum output of the critic with respect to the action computed by the actor network and $G_{\pi_i}$ is the gradient of the actor output with respect to the actor parameters.
Algorithm Flow would be:

1. Initialize the actor and critic networks.

2. Sample a batch of transitions from the replay buffer.

3. Compute the target value functions using the target critic networks and the target policy.

4. Update the critic networks by minimizing the loss function.

5. Update the actor network using the policy gradient.

6. Update the target networks with delayed updates.

7. Repeat steps 2-6 until convergence or a desired number of iterations.

\subsection {DRL Applied to LCC-S Converter Control}
\hspace{10pt} To demonstrate the effectiveness of the TD3 algorithm in tuning the PI controller of an LCC-S converter, it was implemented in MATLAB/SIMULINK. The state vector for the agent was defined as:
\begin{equation}
\mathbf{x} = \left[ \int e(t) . dt \quad e(t) \right]^T
\end{equation}

where, $e(t)=V_{ref}(t) - V_{out}(t)$ is the voltage error. The reward function was designed to motivate the agent to minimize the voltage error:
\begin{equation}
R = - (V_{ref}(t) - V_{out}(t))^2 
\end{equation}
\hspace{10pt} To encourage exploration and improve robustness, the reference voltage, $V_{ref}$, was randomly selected within the range of 190 V to 210 V. This range was chosen due to limitations on adjusting the frequency, as suggested by WPT standards \cite{SAEJ2954_2019}, to regulate the output voltage. By focusing on voltages between 190 V and 210 V, we ensured the system remained within the standard frequency range, preventing any exceedance. The initial values for the PI gains, $K_p$ and $K_i$, were set to 5.28 and 0.05, respectively.

\hspace{10pt} A fully connected two-layer neural network (FCNN) was used to parameterize the PI controller. The actor network output layer consisted of two neurons representing the $K_p$ and $K_i$ gains. Two critic networks were utilized to evaluate the action-value function. Each critic network consisted of a state input layer, an action input layer, and the following 3-layer FCNN within relu-type activation functions. The final layer produced a single-scalar output representing the estimated Q-value.

\hspace{10pt} The learning rate for both actor and critic networks was established at $1 \times 10^{-4}$, a common choice in deep reinforcement learning applications. A total of 500 training episodes were conducted to allow the agent to learn and adapt its behavior over time. To enhance stability and prevent premature convergence, small random noise with a standard deviation of $\sqrt{0.1}$ was added to the target actions. This technique, known as target policy smoothing, helps reduce the risk of the agent becoming trapped in suboptimal local minima.

\hspace{10pt} By incorporating these configurations into the TD3 algorithm, the agent effectively tuned the PI controller, as illustrated in Fig. 8. After 500 episodes, the learning process converged to a maximum reward, resulting in optimized PI controller parameters: $K_p=0.0553$ and $K_i=12.9637$. The reward curves in Fig. 8 illustrate the comparative performance of the TD3 and DDPG algorithms over 500 episodes. The DDPG algorithm demonstrates a sharp initial increase in reward, reflecting rapid early learning. However, its trajectory exhibits significant fluctuations and instability during intermediate episodes, which eventually resulted in an average reward of approximately $-2.5 \times 10^6$. In contrast, the TD3 algorithm achieves a smoother learning curve, stabilizing later and steadily improving to a final reward of approximately $-1.0 \times 10^6$. In particular, the critic and actor parameters for both algorithms were configured similarly, highlighting that the superior performance of TD3 stems from its architectural enhancements, including twin Q-networks to mitigate overestimation bias and delayed policy updates. These features enable TD3 to exhibit greater stability and robust convergence compared to DDPG.

\hspace{10pt} As Fig. 9 depicts, the simulation results highlight the superior performance of the TD3 tuned PI controller in regulating the output voltage of the LCC-S resonant converter compared to the DDPG tuned controller. At $t = 0~\text{s}$, when the step input is triggered to regulate the output voltage to 200 volts, the TD3 tuned controller demonstrates reduced overshoot and better stability, ensuring a smoother response. At $t = 0.04~\text{s}$, following a drop in 5\% in input voltage and at $t = 0.07~\text{s}$, after a drop in load 10\%, the TD3-tuned controller maintains consistent output regulation with less oscillations compared to the DDPG-tuned controller. Throughout the simulation, under the influence of random noise at 85 kHz (matching the switching frequency), the TD3-tuned controller exhibits enhanced robustness by reducing voltage ripple, making it more reliable for real-world applications despite not being faster in response. This comparison underscores TD3's ability to deliver more stable and resilient control performance under dynamic and noisy conditions.

\begin{figure}[t]
  \begin{center}
  \includegraphics[width=3.5in]{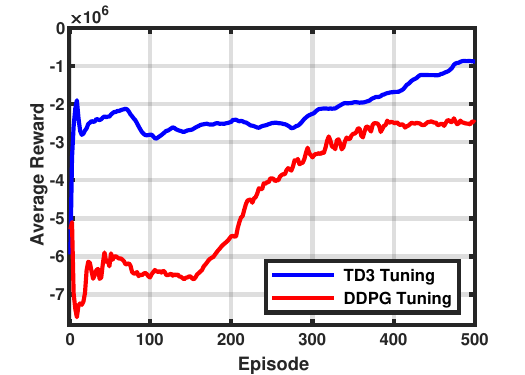}\\
  \vspace{-8pt}
  \caption{Average reward convergence in DRL-based PI tuning.}\label{Average_Reward_DRL}
  \end{center}
\end{figure}


\begin{figure}[t]
  \begin{center}
  \includegraphics[width=3.5in]{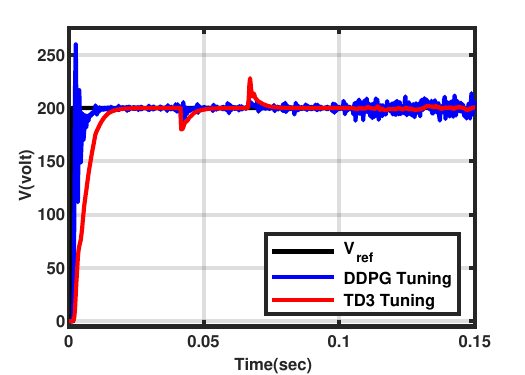}\\
  \vspace{-8pt}
  \caption{Step response of the LCC-S converter with random PI parameters vs. TD3-tuned PI parameters.}\label{Average_Reward_DRL}
  \end{center}
\end{figure}

\section{Experimental Result}
\hspace{10pt} The experimental setup (Fig. 10) was conducted in the Photovoltaic Lab at the University of Tehran and includes the WPT system featuring an LCC-S resonant converter. The setup consists of primary and secondary LCC-S converter boards, circular magnetic structure pads for magnetic coupling, a flyback converter providing the converter’s input voltage, and a high-power resistor used as the converter’s load. A buck-boost converter supplies four distinct power rails for gate drivers, while an STM32F407VGT ARM Cortex-M microcontroller manages the control loop by adjusting the switching frequency to maintain output voltage regulation. The system operates with an input voltage of 28 V and a 45 ohm output resistor. A 12-bit ADC with Direct Memory Access (DMA) handles output voltage sampling, and a complementary 50\% duty cycle PWM controls the switching frequency, constrained by a saturation block between 79 and 90 kHz.
\begin{figure}[t]
  \begin{center}
  \includegraphics[width=2.8in]{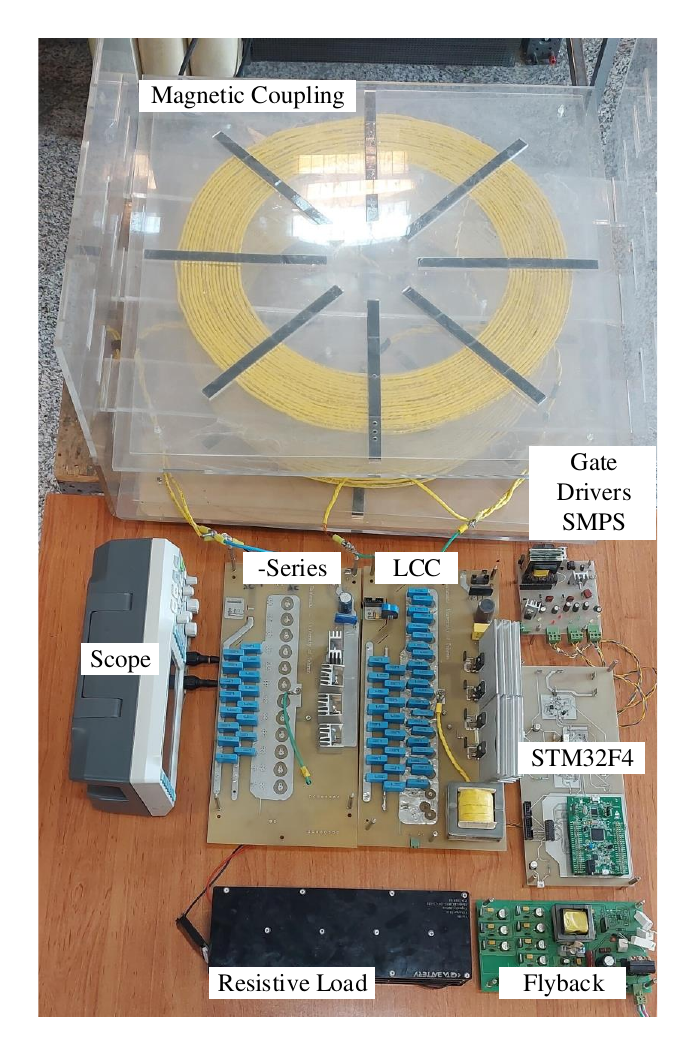}\\
  \vspace{-15pt}
  \caption{Experimental setup of the WPT system in the photovoltaic lab at the University of Tehran.}\label{Average_Reward_DRL}
  \end{center}
\end{figure}

\hspace{10pt} Fig. 11 presents oscilloscope waveforms of the magnetic coupling coils (Fig. 11a) and the LCC-S converter's input and output voltages (Fig. 11b). The coil voltages reveal a drop due to air gaps and environmental noise, which leads to waveform distortion. Under closed-loop control, the microcontroller firmware adjusts the output voltage at a pulse frequency of one hertz, maintaining the target of 28 V with minimal overshoot (under 2\%), a 5 ms rise time, and an average output voltage of 27.49 V, indicating a 2\% error. The input voltage shows distortion during this closed-loop operation but remains stable, thanks to the preceding flyback converter stage.

\begin{figure}[t]
  \begin{center}
  \includegraphics[width=3.3in]{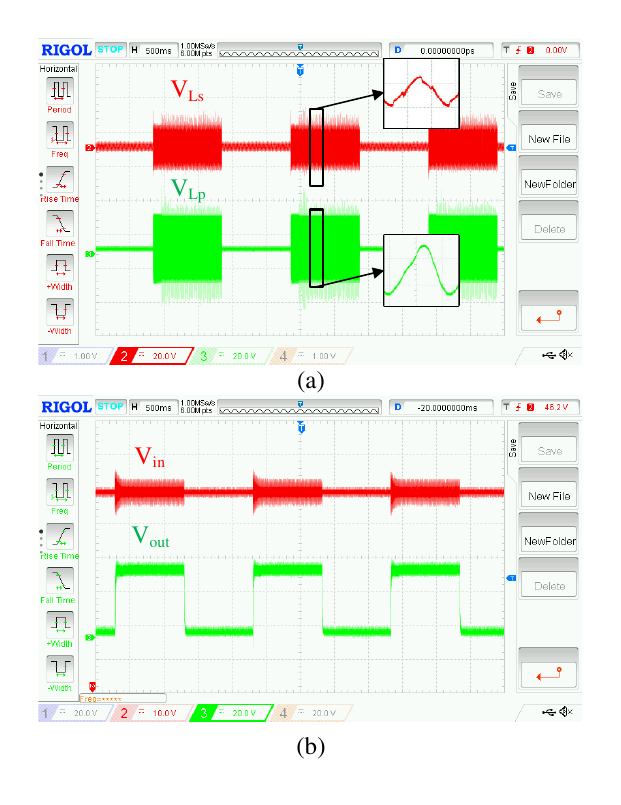}\\
  \vspace{-22pt}
  \caption{Oscilloscope waveforms of coupling coil voltages (a) and LCC-S converter input/output voltages (b).}\label{Average_Reward_DRL}
  \end{center}
\end{figure}

\section{Conclusion}
\hspace{10pt} In this paper, we proposed a novel DRL-based frequency control strategy for LCC-S resonant converters, addressing key limitations in traditional control methods. Unlike existing approaches that rely on manually tuned PI controllers, we introduced an automated tuning framework using the TD3 algorithm, enabling adaptive optimization of controller parameters. The converter was modeled using the DPWA method, identifying six subsystems to capture its dynamic behavior with more accuracy. The simulation results highlighted the challenges posed by the converter’s non-linearities, which conventional controllers struggle to handle. By leveraging DRL, our approach allowed the PI controller to iteratively learn optimal parameters, significantly improving robustness and control performance. Both simulation and experimental results demonstrated the effectiveness of DRL-based tuning, confirming its potential to enhance robust control in complex power electronic systems.


%





\ifCLASSOPTIONcaptionsoff
  \newpage
\fi





\bibliographystyle{IEEEtran}
\bibliography{IEEEabrv,Bibliography}
%

\vspace{-25pt}
\begin{IEEEbiography}[{\includegraphics[width=1in,height=1.25in,clip,keepaspectratio]{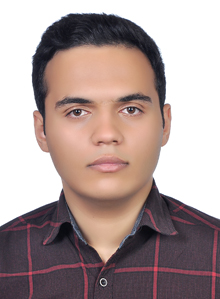}}]{Reza Safari}
was born in 1997 in Iran. He received a B.Sc. degree in electrical engineering (control and systems) from K. N. Toosi University of Technology, Tehran, Iran, in 2019, and the M.Sc. degree in electrical engineering (power electronics) from the University of Tehran, Tehran, Iran, in 2022. His work focuses on embedded systems and software development for power electronics and automotive applications. His research interests include wireless power transmission, power electronics control theory, microgrid control, and converter modeling.
\end{IEEEbiography}
\vspace{-25pt}
\begin{IEEEbiography}[{\includegraphics[width=1in,height=1.25in,clip,keepaspectratio]{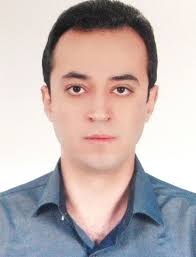}}]{Mohsen Hamzeh}
(S’09–M’13) received the B.Sc. and M.Sc. degrees from the University of Tehran, Tehran, Iran, in 2006 and 2008, respectively, and the Ph.D. degree from Sharif University of Technology, Tehran, Iran, in 2012, all in electrical engineering. From 2013 to 2018, he was an Assistant Professor at Shahid Beheshti University, Tehran, Iran. In 2018, he joined the School of Electrical and Computer Engineering, University of Tehran. His research interests include renewable energies, microgrid control and applications of power electronics in power distribution systems.
\end{IEEEbiography}
\vspace{-25pt}
\begin{IEEEbiography}[{\includegraphics[width=1.1in,height=1.30in,clip,keepaspectratio]{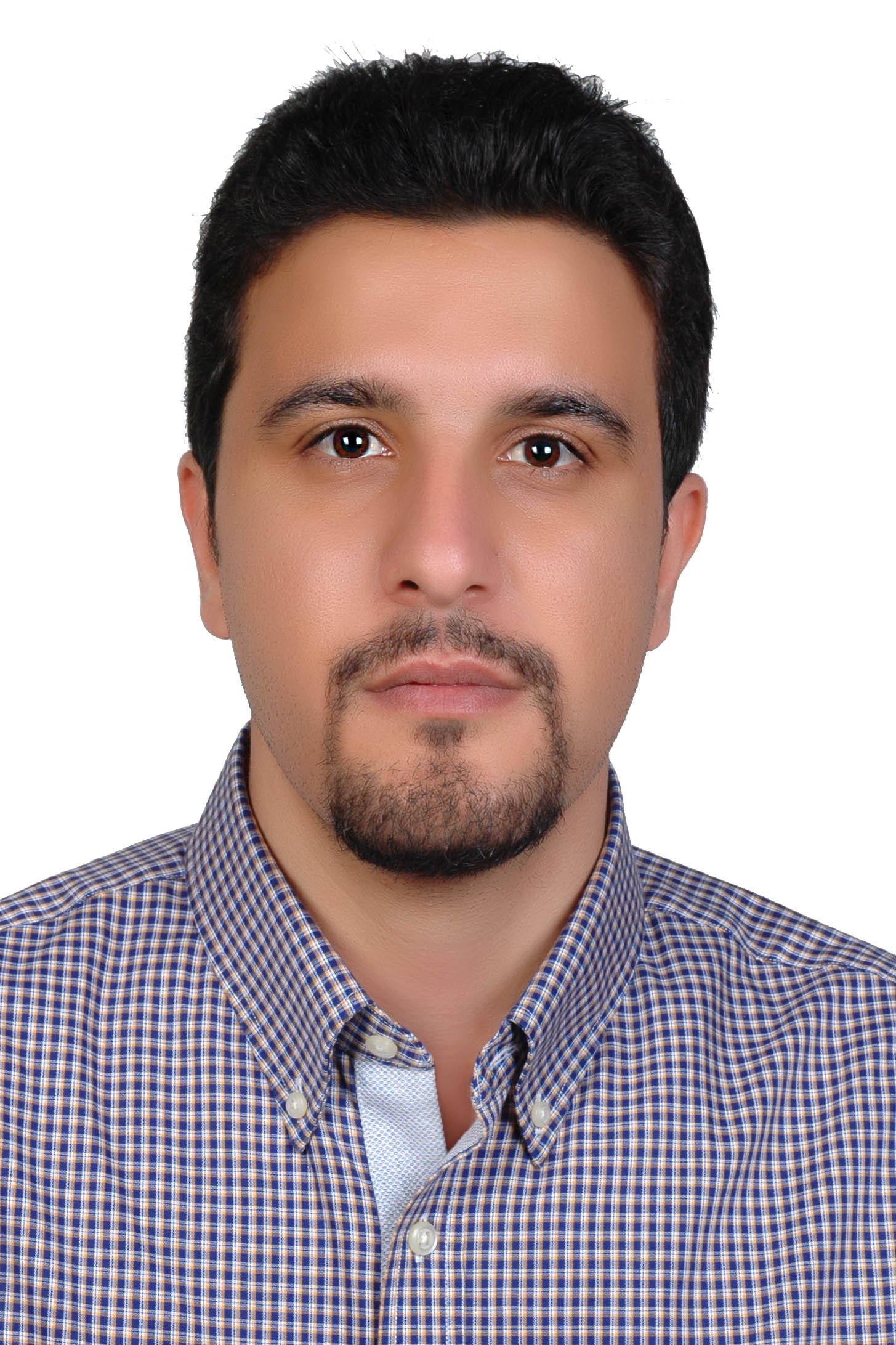}}]{Nima Mahdian Dehkordi}
  received M.Sc. and Ph.D. degrees from Sharif University of Technology, Tehran, Iran, in 2012 and 2016, respectively, both in electrical engineering.  From 2016 to 2019, he was an Assistant Professor in the Department of Electrical Engineering, Science and Research Branch, Islamic Azad University, Tehran, Iran.  He joined the Department of Electrical Engineering, Shahid Rajaee Teacher Training University, Tehran, Iran, in 2019, where he is currently an Associate Professor. His research interests include control systems, applications of control theory in power electronics, microgrid control, distributed and cooperative control, the Internet of Things, nonlinear control, and network control. 
\end{IEEEbiography}





\vfill


\end{document}